\documentclass[runningheads,a4paper]{llncs}

\usepackage{amssymb}
\usepackage{caption}
\usepackage{graphicx,longtable,wrapfig}
\usepackage{amsmath}
\usepackage{epstopdf}
\usepackage[ruled,vlined]{algorithm2e}

\usepackage{url}
\newcommand{\keywords}[1]{\par\addvspace\baselineskip
\noindent\keywordname\enspace\ignorespaces#1}

\begin{document}
\newcommand{\E}{\mathsf{E}}
\newcommand{\Prob}{\mathsf{P}}
\newcommand{\G}{\mathsf{\Gamma}}
\newcommand{\BA}{Barab\'asi--Albert}
\newcommand{\BO}{Buckley--Osthus}
\newcommand{\HK}{Holme and Kim}

\mainmatter  

\title{Evolution of the Media Web}

\titlerunning{Evolution of the Media Web}

%
%
\author{Damien~Lefortier
\and Liudmila~Ostroumova \and Egor~Samosvat}
\authorrunning{Lefortier D., Ostroumova L., Samosvat E.}
\urldef{\mailsa}\path|{damien, ostroumova-la, sameg}@yandex-team.ru|

\institute{Yandex, Moscow, Russia\\
\mailsa\\
}

%
%

\maketitle

\begin{abstract}

  We present a detailed study of the part of the Web related to media
  content, i.e., the Media Web. Using publicly available data, we
  analyze the evolution of incoming and outgoing links from and to
  media pages. Based on our observations, we propose a new class of
  models for the appearance of new media content on the Web where
  different \textit{attractiveness} functions of nodes are possible
  including ones taken from well-known preferential attachment and
  fitness models.  We analyze these models theoretically and
  empirically and show which ones realistically predict both the
  incoming degree distribution and the so-called \textit{recency property} of
  the Media Web, something that existing models did not do well.
  Finally we compare these models by estimating the likelihood of the
  real-world link graph from our data set given each model and obtain
  that models we introduce are significantly more likely than
  previously proposed ones. One of the most surprising results is that
  in the Media Web the probability for a post to be cited is
  determined, most likely, by its quality rather than by its
  current popularity.

\keywords{Media Web, random graph models, recency.}

\end{abstract}

\section{Introduction}\label{Introduction}

Numerous models have been suggested to reflect and predict the growth
of the Web \cite{Boccaletti06,Bonato04,Kumar00}. The most well-known
ones are preferential attachment models (see Section~\ref{Related_work} for a
more thorough discussion about previous work). One of the main
drawbacks of these models is that they pay too much attention to old
pages and do not realistically explain how links pointing to
newly-created pages appear (as we discuss below). In this paper, we
are interested in the Media Web, i.e., the highly dynamic part of the Web related to media content where
a lot of new pages appear daily. We show that the Media Web has some specific properties and should therefore be analyzed separately. Note that some other parts of the Web have already been studied, for example in \cite{Leskovec08} a model for the Social Web is suggested.

Most new media pages like news and blog posts are popular only
for a short period of time, i.e., such pages are mostly cited and
visited for several days after they appeared \cite{2013timely}.  We analyze this
thoroughly later in the paper and introduce a \textit{recency
property}, which reflects the fact that new media pages tend to connect
to other media pages of similar age (see Section~\ref{Motivation}).

In this context, we propose a new class of models for the appearance
of new media content on the Web where different
\textit{attractiveness} functions of nodes are possible including ones
taken from well-known preferential attachment and fitness models, but
also new ones accounting for specificities of the Media Web.  We
analyze these models theoretically and empirically using MemeTracker
public data set \cite{MemeData} and show which ones realistically
predict both the incoming degree distribution and the \textit{recency property}
of the Media Web, something that existing models did not do well.
Finally we compare these models by estimating the likelihood of the
real-world link graph from this data set given each model and obtain
that models we introduce in this paper are significantly more likely
than previously proposed ones.  One of the most surprising results is
that in the Media Web the probability for a post to be cited is
determined, most likely, by its quality rather than by its current
popularity.

The contributions of this paper are the following:
\begin{itemize}
\item We suggest a new class of models for the appearance of new media
  content on the Web where different
  \textit{attractiveness} functions of nodes are possible;
\item We analyze these models
  theoretically and empirically and show which ones realistically
  depict the behavior of the Media Web;
\item We compare these models by
  estimating the likelihood of the real-world link graph from our data set given each model.
\end{itemize}

The rest of the paper is organized as follows. In Sections \ref{Related_work} and \ref{Motivation}, we discuss related work and experimental results, which motivated this
work. In Section~\ref{Model}, based on the results of our experiments, we define our class of
models. We analyze theoretically some properties of these models in Section~\ref{Theory}, while in Section~\ref{MLE} we validate our models by computing the likelihood of the real-world link graph from our data given each model.

\section{Related Work}\label{Related_work}

One of the first attempts to propose a realistic mathematical model of
the Web growth was made in \cite{Barabasi99}. The main idea is to
take into account the assumption that new pages often link
to old popular pages. Barab\'{a}si and Albert defined a graph
construction stochastic process, which is a Markov chain of graphs,
governed by the \emph{preferential attachment}. At each step in
the process, a new node is added to the graph and is joined to $m$
different nodes already existing in the graph that are chosen with
probabilities proportional to their incoming degree (the measure of
popularity). This model successfully explained some properties of the
Web graph like its small diameter and power law incoming degree
distribution. Later, many modifications to the \BA~model have been
proposed, e.g., \cite{Buckley04,Holme02,Cooper03}, in order to more
accurately depict these but also other properties (see
\cite{Albert02,Bollobas03} for details).

It was noted by Bianconi and Barab\'{a}si in \cite{Bianconi01} that in
real networks some nodes are gaining new incoming links not only because of
their incoming degree, but also because of their own intrinsic properties. For
example, new Web pages containing some really popular content can acquire a
large number of incoming links in a short period of time and become
more popular than older pages. Motivated by this observation, Bianconi
and Barab\'{a}si extended preferential attachment models with pages'
inherent quality or \textit{fitness} of nodes.
When a new node is added to the graph, it is joined to some already existing
nodes that are chosen with probabilities proportional to the product
of their fitness and incoming degree. This model was theoretically analyzed in \cite{Borgs07}.

In the context of our research, the main drawback of these models is
that, as said, they pay too much attention to old pages and do not
realistically explain how links pointing to newly-created pages
appear. Note also that highly dynamic parts of the Web like social
networks or weblogs exhibit a specific behavior and should therefore
be modeled separately (see Section~\ref{Motivation}).  In \cite{Leskovec08}, the evolution of social
networks, or the Social Web, was thoroughly investigated and, based on their results, a
model was suggested. In turn, we suggest a model for the Media Web.  The main idea is
to combine preferential attachment and fitness models with a recency
factor. This means that pages are gaining incoming links according to
their \textit{attractiveness}, which is determined by the incoming
degree of the page, its \textit{inherent popularity} (some
page-specific constant) and age (new pages are gaining new links more
rapidly).

\section{Recency Property of the Media Web}\label{Motivation}

In this section, we present experiments, which motivated us to propose
a new model for the Media Web. Our model is based on these experimental results.

\subsection{Experimental Setup}\label{Setup}

We use MemeTracker public data set \cite{MemeData}, which covers 9
months of Media Web activity -- quite a significant time period.  Note
that only outgoing links from the content
part of the post were extracted (no toolbar, sidebar links).  See
\cite{Leskovec09} for details on how this data was collected.

From this data set we kept only links pointing to documents also in
the data set, i.e., links with known timestamps both for the source and
the destination. We assume that these timestamps correspond to the
time when each document was posted on the Web, and we also filtered
out links for which the timestamp of the destination is greater than
for the source (impossible situation). This can happen because timestamps are noisy and therefore not always
reliable. We finally obtained a data set of about 18M links and 6.5M documents
that we use in the following experiments.


\subsection{Recency Property}\label{Recency_experiments}

\begin{figure}
        \centering
        \includegraphics[width=0.68\textwidth]{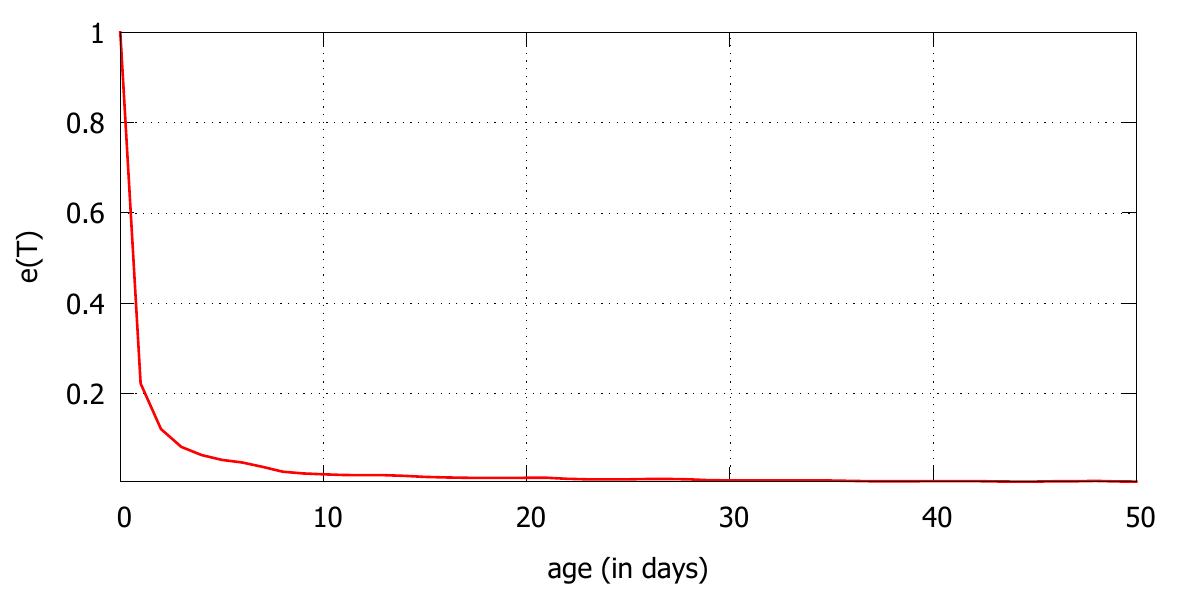}
        \caption{The recency property}
        \label{fig:recency_real}
\end{figure}

Let us define the \textit{recency property} for a graph evolving in time.
Denote by $e(T)$ the fraction of edges connecting nodes whose age difference is greater than $T$.
We analyze the behavior of $e(T)$ and show that media pages tend to connect to pages of similar age.
We plotted $e(T)$ for our dataset and noted that $e(T)$ is decreasing exponentially fast (see Figure~\ref{fig:recency_real}), which is not the case for preferential attachment model as we show later in this paper (Section~\ref{Recency}).

\section{Model}\label{Model}

Suppose that we have a fixed set of hosts $H_1, \dots, H_n$.
Each host $H_i$ has its own rate of new pages appearance $\lambda_i$.
At the beginning of the process, we have no pages.
We assume that new pages appear on a host $H_i$ according to a Poisson process with parameter $\lambda_i$.
A Poisson process is often used to model a sequence of random events that happen independently with a fixed rate over time\footnote{\url{http://en.wikipedia.org/wiki/Poisson_process}}.
Poisson processes for different hosts are independent.

When a new page $p$ is created on a host $i$, it has $m_p$ mutually independent
outgoing links pointing to already existing media
pages. The target page of each link is determined as follows. First, the target host $k$ is chosen with probability $\rho_{ik}$ ($\sum_{k=1}^n \rho_{ik} = 1$). Then, the probability of each page $r$ on the host $k$ to be chosen is
proportional to the \emph{attractiveness} of $r$, which is some function
of $d_r$ (current incoming degree of $r$), $q_r$ (intrinsic quality of $r$),
and $a_r$ (current age of $r$). Different attractiveness functions are
possible:

$$f_{\tau_k}(d, q, a) =
  \left(1 \text{ or } q \right) \cdot
  \left(1 \text{ or } d \right) \cdot
  \left(1 \text{ or } e^{-\frac{a}{\tau_k}}  \right) \,.$$
Where $\tau_k$ corresponds to the mean lifetime of the decaying attractiveness for media pages on host $k$.

E.g., $f_{\tau_k}(d, q, a) = d$ leads to preferential
attachment, while $f_{\tau_k}(d, q, a) = q \cdot d$ leads to
fitness model. In this paper, we study different options and show
which ones best depict the behavior of the Media Web.

Let us denote by $\Omega(H_i)$ the set of pages, which belong to a host $H_i$.
We assume that the distributions of $q_p$ and $m_p$ for $p \in \Omega(H_i)$
are the properties of $H_i$. The only thing we assume about these
distributions is that $q_p$ and $m_p$ have finite expectations.

\section{Theoretical analysis}\label{Theory}

\subsection{Incoming degree distribution}\label{Degree}

In \cite{Bianconi01,Bollobas01,Buckley04}, models without \emph{recency
factor} (i.e., without the factor $e^{-\frac{a}{\tau_k}}$ in the
attractiveness function) have been analyzed.  On the contrary, in this paper we show that we need the recency factor to reflect
some important properties of the Media Web (see Section~\ref{Recency}).  Therefore we assume here that
the attractiveness function has such recency factor.

Denote by $d_p(q_p, t, t_p)$ the incoming degree at
time~$t$ of a page $p$ created at time $t_p$ with intrinsic
quality $q_p$.  Let us also define,  for each host $H_k$, the average attractiveness of its pages at time $t$:
\begin{equation}\label{W_k}
W_k(t) = \mathbb{E} \sum_{p\in\Omega(H_k)} f_{\tau_k}(d_p(q_p, t, t_p), q_p, t - t_p)\,.
\end{equation}
We will show in this section that $W_k(t) \to W_k$ as $t \to \infty$, where $W_k$ are some positive constants.

Let $M_k$ be the average number of outgoing links of pages $p \in \Omega(H_k)$.
Then $N_k = \sum_i \lambda_i M_i \rho_{ik}$ is the average rate of new links
pointing to host $H_k$ appearance.

\begin{theorem}\label{Theorem1}
Let $p \in \Omega(S_k)$ be a page with quality $q_p$ and time of creation $t_p$.
\begin{itemize}
\item[(1)] If $f_{\tau_k}=q \cdot d \cdot e^{-\frac{a}{\tau_k}}$, \\ then
  ${d_p(q_p, t, t_p) = e^{\frac{N_k \tau_k q_p}{W_k}\left(1 - e^{\frac{t_p-t}{\tau_k}} \right)}}$,\\
\item[(2)] If $f_{\tau_k}=q \cdot e^{-\frac{a}{\tau_k}}$, \\ then $d_p(q_p, t, t_p) = \frac{N_k \tau_k q_p}{W_k}\left(1 - e^{\frac{t_p-t}{\tau_k}} \right)$.
\end{itemize}
\end{theorem}

It follows from Theorem \ref{Theorem1} that in the first case, in
order to have a power law distribution of $d_p$, we need to have $q_p$
distributed exponentially. In this case, for each host, the parameter of
the power law distribution equals $\frac{N_k\tau_k\mu}{W_k}$, where $\mu$
is the parameter of exponential distribution. It is interesting to note
that this latter parameter cannot affect the
parameter of the power law distribution. Indeed, if we multiply $\mu$ by some
constant, then $W_k$ will also be multiplied by the same
constant (see (\ref{W_k})). Therefore, we can change the parameter of the power law
distribution only by varying $N_k$ and $\tau_k$. The problem is that the constant $W_k$ depends on $N_k$ and $\tau_k$ (see equation~(\ref{W}) in the proof). Hence, it is impossible to find analytical expressions for $N_k$ and $\tau_k$, which give us the desired parameter of the power law distribution.

In the second case, a power law distribution of $q_p$ leads to a power
law distribution of $d_p$ with the same constant. Therefore, it is easy to get
a realistic incoming degree distribution in this case.

In both cases, we cannot avoid the quality factor because if we do not have it
in the attractiveness function (i.e., if $q_p$ is constant for all
media pages), then the solution does not depend on $q_p$ and we do not have a
power law for the incoming degree distribution.

\begin{figure}
        \centering
        \includegraphics[width=0.68\textwidth]{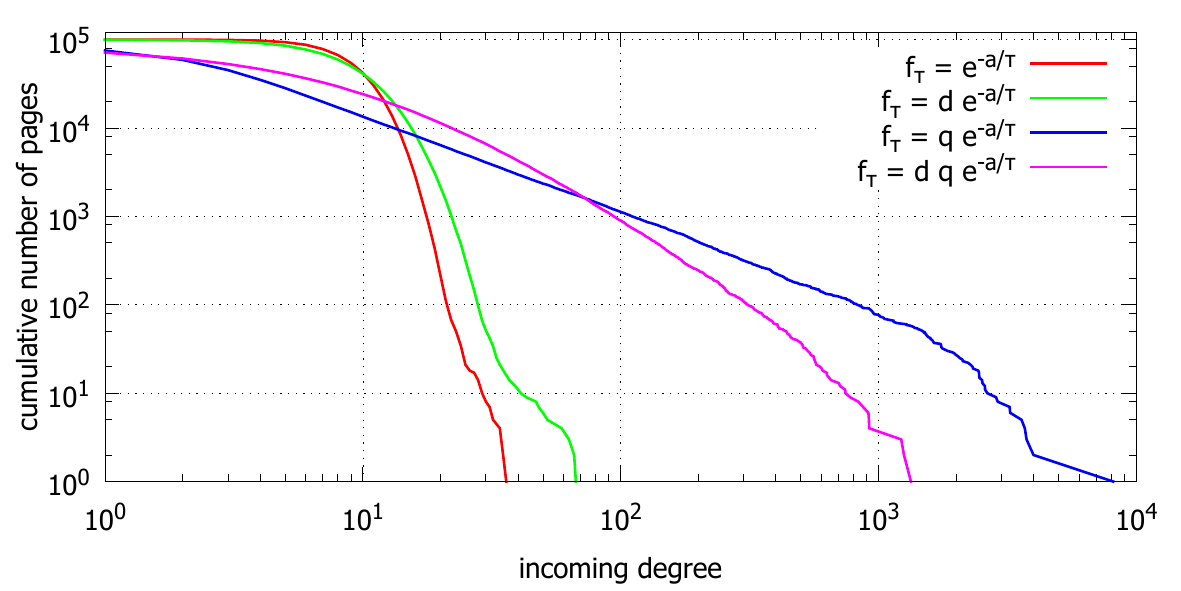}
        \caption{Incoming degree distribution for each model}
        \label{fig:distr_model}
\end{figure}

To illustrate the results of Theorem \ref{Theorem1}, we generated
graphs according to our model with different functions
$f_{\tau_k}$. Obtained results are shown on
Figure~\ref{fig:distr_model}.

\begin{proof}

In mean-field approximation, we have the following differential
equation:
$$
\frac{\partial d_p(q_p, t, t_p)}{\partial t} =
N_k \frac{f_{\tau_p}(d_p(q_p, t, t_p), q_p, t - t_p)}
       {W_k(t)},
$$
here $p \in \Omega(H_k)$.


In the case $f_{\tau_k}(d, q, a) = q \cdot d \cdot e^{-\frac{a}{\tau_k}}$ we have:

\begin{equation}\label{MeanField}
\frac{\partial d_p(q_p, t, t_p)}{\partial t} =
N_k \frac{q_p \cdot d_p(q_p, t, t_p) \cdot e^{-\frac{t-t_p}{\tau_k}}}
       {W_k(t)}
\end{equation}

Later in this section, we show that for each $k$, $W_k(t)$ tends to some
positive constant $W_k$: $\lim_{t \rightarrow \infty} W_k(t) =
W_k$.

We thus have the following solution of the equation
(\ref{MeanField}):

$$
d_p = e^{ \frac{N_k \tau_k q_p}{W_k} \left(1 - e^{\frac{t_p-t}{\tau_k}} \right) } \xrightarrow{\scriptscriptstyle t\to\infty} e^{ \frac{N_k \tau_k q_p}{W_k} }
$$

In case $f_{\tau_k}(d, q, a) = q \cdot e^{-\frac{a}{\tau_k}}$,
by similar but even simpler calculations, we obtain:

$$
d_p = \frac{N_k \tau_k q_p}{W_k} \left(1 - e^{\frac{t_p-t}{\tau_k}} \right) \xrightarrow{\scriptscriptstyle t\to\infty}  \frac{N_k \tau_k q_p}{W_k}
$$


Let us now check that $\lim_{t \rightarrow \infty} W_k(t)$ is indeed a constant.
Consider the case $f_{\tau_k}(d, q, a) = q \cdot d \cdot e^{-\frac{a}{\tau_k}}$.
Let $\rho_k(q)$ be the probability density function of $q_p$ for $p \in \Omega(H_k)$. Therefore:
\begin{multline*}
W_k(t) = \int_0^{\infty} \left( \int_0^t \lambda_k  q \rho_k(q) d(q, t, x) \cdot e^{-\frac{t-x}{\tau_k}} d x \right) d q = \\
=  \int_0^{\infty}
  \left( \int_0^t  \lambda_k q \rho_k(q)
     e^{ \frac{N_k \tau_k q}{W_k} \left(1 - e^{\frac{x-t}{\tau_k}} \right) }
     \cdot e^{\frac{x-t}{\tau_k}} d x
  \right) d q = \\
= \int_0^{\infty} \frac{\lambda_k W_k}{N_k} \left( e^{ \frac{N_k \tau_k q}{S} \left(1 - e^{\frac{-t}{\tau_k}}\right) } - 1 \right) \rho_k(q) d q \, .
\end{multline*}
Thus for $W_k$ we finally have the following equation:
\begin{equation}\label{W}
W_k = \lim_{t \rightarrow \infty} W_k(t) =
\underbrace{\frac{\lambda_k W_k}{N_k} \left( \int_0^{\infty}  e^{ \frac{N_k \tau_k q}{W_k} } \rho_k(q) d q - 1 \right)}_{F_k(W_k)}.
\end{equation}
There is a unique solution of the equation (\ref{W}). To show this, we first check that ${y = F_k(x)}$ is monotone:
$$
F_k'(x) = \frac{\lambda_k}{N_k} \left( \int_0^{\infty}  e^{ \frac{N_k \tau_k q}{x}} \left(1- \frac{N_k \tau_k q}{x}\right) \rho_k(q) d q - 1 \right) \le 0\, ,
$$
since:
$$
e^{ \frac{N_k \tau_k q}{x}} \left(1- \frac{N_k \tau_k q}{x}\right) \le 1 \text{ and } \int_0^{\infty}  \rho_k(q) d q = 1.
$$
Also ${F_k(x) \to \tau_k \lambda_k \mathbb{E}_{p \in \Omega(S_k)} q_p}$ as ${x \to \infty}$
and $F_k(x) \to \infty$ as $x \to 0$. From these observations, it follows that
$y=x$ and $y=F_k(x)$ have a unique intersection. In other words, the equation~(\ref{W}) has a unique solution.


Similarly, we
can show that $\lim_{t \rightarrow \infty} W_k(t)=W_k$ for the attractiveness
function $f_{\tau_k}=q \cdot e^{-\frac{a}{\tau_k}}$.




\end{proof}


\subsection{Recency property}\label{Recency}

In this section, we show that we need a recency factor
$e^{-\frac{a}{\tau_k}}$ in the formula for the attractiveness function $f_{\tau_k}$.
We prove that because of the recency factor, the number
of edges, which connect nodes with time difference greater than $T$
decreases exponentially in $T$. We prove the following theorem.

\begin{theorem}\label{Theorem2}
For $f_{\tau_k}=q \cdot d \cdot e^{-\frac{a}{\tau_k}}$ or $f_{\tau_k}=q \cdot e^{-\frac{a}{\tau_k}}$ we have
$$
e(T) \sim  \sum_k N_k C_k e^{\frac{-T}{\tau_k}}\,,
$$
where $C_k$ are some constants.
\end{theorem}



\begin{figure}
        \centering
        \includegraphics[width=0.68\textwidth]{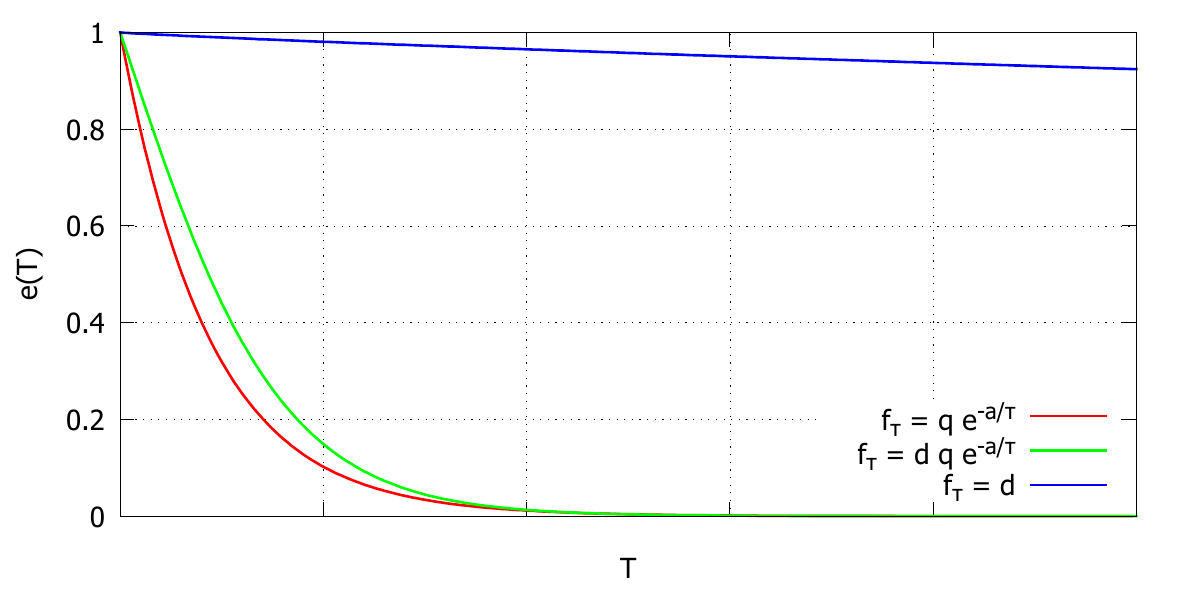}
        \caption{Recency property in the model}
        \label{fig:recency_model}
\end{figure}

Due to space constraints, we move the proof of Theorem~\ref{Theorem2} to Appendix.
To illustrate the results obtained, we plot $e(T)$ for different attractiveness
functions on
Figure~\ref{fig:recency_model}. Note that if we have a recency factor in the attractiveness
function, then $e(T)$ approaches its upper bound exponentially
fast. In contrast, if the attractiveness function equals $d$ (preferential
attachment), then $e(T)$ grows almost linearly with a small rate.

\section{Validation}\label{MLE}

The idea of using Maximum Likelihood in order to compare different graph models and estimate their parameters was suggested in \cite{Bezakova06}. Since then this method was used for several models (see, e.g., \cite{Leskovec08,Leskovec10}). Motivated by these works we also use the idea of Maximum Likelihood in order to compare new models we suggest in this paper with preferential attachment and fitness models.

\subsection{Parameters estimation}

In order to do simulations, we first need to estimate all parameters
of our models. Note that we are not trying to find the best
parameters here. Instead we propose to use simple estimations, which are enough to show the
improvements obtained by using our new models.

\textbf{Host-to-host probabilities.}
We estimated the matrix $\rho_{ij}$ by counting the fraction of edges going from hosts $H_i$ to $H_j$. Note that 74\% of all edges are host internal. We also add host to host probabilities to fitness and preferential attachment models and, as we show later in Section~\ref{likelihood}, this assumption allows to improve these models.

\textbf{Estimation of $\tau$.}
In order to estimate $\tau_k$ for each host $H_k$, we consider the histogram of age difference of connected pages.
Let $x_i$ ($i\ge0$) be the number of links which connect pages with age difference greater than $i$ but less than $i+1$ days. If we assume an exponential decay, then for $i<j$ we have $\frac{x_i}{x_j} = e^{\frac{(i-j)T}{\tau_k}}$, i.e.,  $\tau_k = \frac{(i-j)T}{\log{\frac{x_i}{x_j}}}$, where $T$ is the time interval of one day. Therefore, we take:
$$
\tau_k = \sum_{\substack{0\le i< j< 10: \\ x_i \neq 0, x_j \neq 0}} \frac{(i-j)T}{{10\choose 2}\log{\frac{x_i}{x_j}}}.
$$

We make a cut-off at 10 days because even though the tail of the histogram is heavier than exponential, the most important for us is to have a good estimation when pages are young, i.e. when most incoming links appear.

\textbf{Estimation of quality.}
Given the final incoming degree $d$ of a node, we can use Theorem~\ref{Theorem1} to find its quality,
i.e., we have $q = \frac{W d}{N_k \tau_k}$ in the case of $f_{\tau} = q e^{\frac{-a}{T}}$ and $q = \frac{W \ln d }{N_k \tau_k}$ in the case of $f_{\tau} = d q e^{\frac{-a}{T}}$. Note that the factor $\frac{W}{N_k \tau_k}$ is common for all pages created on host $H_k$ and can be cancelled so we finally used the following estimations: $q = d$ and $q = \ln d$ respectively.

\subsection{Likelihood}\label{likelihood}

In order to valid our model, we propose to use the data described in Section~\ref{Setup} and estimate the likelihood of the real-world link graph from this data set given each model discussed in this paper. We do this as follows.

We add edges one by one according to their historical order and compute
their probability given the model under consideration.
The sum of logarithms of all obtained probabilities gives us the log-likelihood of our graph.
We normalize this sum by the number of edges and obtained results are presented in Table~\ref{tab:log-likelihood}.

\begin{table}
\begin{center}
\caption{Log-likelihood table: average logarithm of edge probability.}\label{tab:log-likelihood}
\begin{tabular}{|c|c|c|c|c|c|c|}
\hline
$d$ & $q$ &  $e^{\frac{-a}{\tau}}$ & $d q$ & $d e^{\frac{-a}{\tau}}$  & $q e^{\frac{-a}{\tau}}$ & $d q e^{\frac{-a}{\tau}}$ \\
\hline
-6.11 & -5.56 & -5.34 & -6.08 & -5.50 & \textbf{-5.17} & -5.45 \\
\hline
\end{tabular}
\end{center}
\end{table}

\begin{figure}
        \centering
        \includegraphics[width=0.66\textwidth]{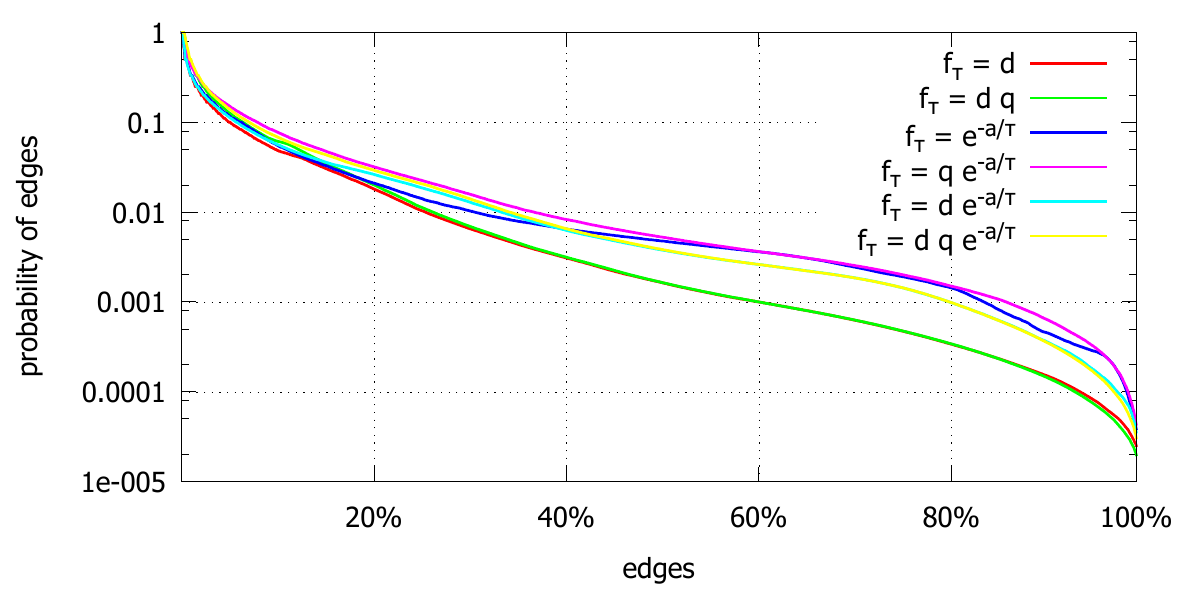}
        \caption{Distribution of edges' probabilities}
        \label{fig:edges}

        \centering
        \includegraphics[width=0.65\textwidth]{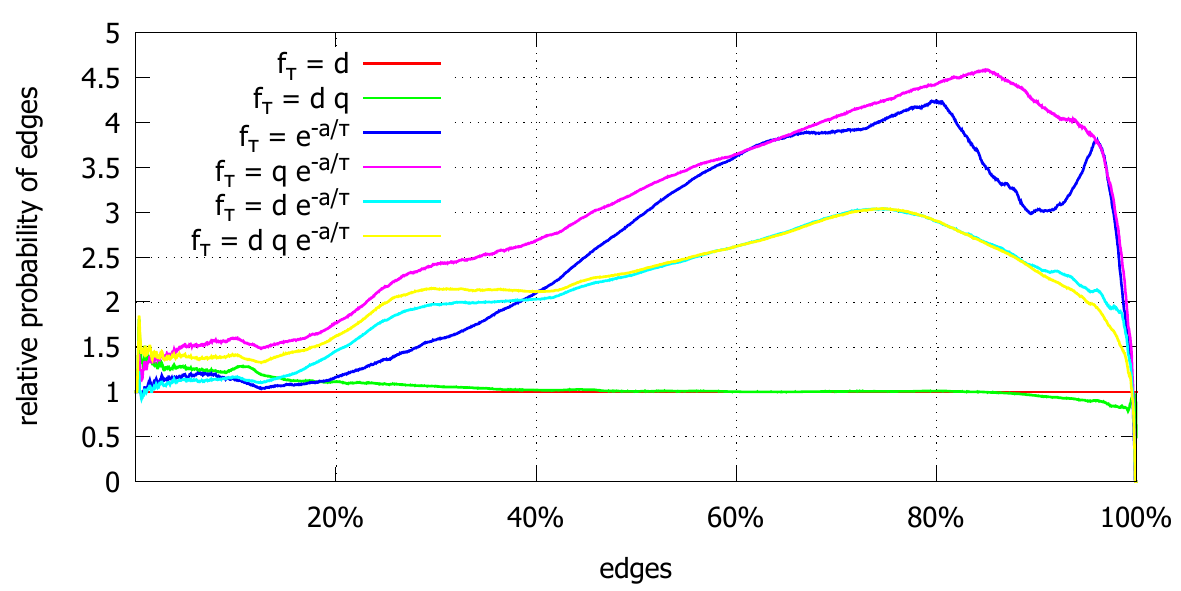}
        \caption{Distribution of edges' probabilities relative to preferential attachment model}
        \label{fig:edges_rel}
\end{figure}

We see that the most likely model here is with $f_{\tau} = q
e^{\frac{-a}{\tau}}$. However, since timestamps are noisy and
therefore not always reliable (see Section~\ref{Setup}), these results
might not be representative (for example, if the probability of one
edge is very small, it can heavily affect the final
likelihood). Hence, in addition to this log-likelihood, which is
strongly affected by outliers, we also performed the analysis of
edges' probabilities, i.e. we try to understand which model is better
on a per-edge basis.
We believe that such deeper analysis allows to reduce the
influence of outliers when validating our models.
To the best of our knowledge, this is the first time such analysis is
made when using Maximum Likelihood in order to compare different graph
models.

Each edge has different probabilities according to different models and there is one model $M$ for which this probability is the largest. In this case, we say that the model $M$ \textit{wins} on this edge (see Table~\ref{tab:winner}). Also, for each pair of models $M_1$ and $M_2$, we computed the percentage of edges which have greater probability according to $M_1$ than according to $M_2$ (see Table~\ref{tab:competition}). It can be clearly seen from both tables that the recency factor plays a very important role.

\begin{table}
\begin{center}
\caption{Winner table: fraction of edges on which model wins all others.}\label{tab:winner}
\begin{tabular}{|c|c|c|c|c|c|c|}
\hline
$d$ & $q$ &  $e^{\frac{-a}{\tau}}$ & $d q$ & $d e^{\frac{-a}{\tau}}$  & $q e^{\frac{-a}{\tau}}$ & $d q e^{\frac{-a}{\tau}}$ \\
\hline
0.03 & 0.07 & 0.28 & 0.07 & 0.07 & \textbf{0.30} & 0.16 \\
\hline
\end{tabular}
\end{center}
\end{table}

\begin{table}
\begin{center}
\caption{Competition table: the value in (a,b) is the fraction of edges where $a$ wins $b$.}\label{tab:competition}
\begin{tabular}{|c|c|c|c|c|c|c|c|}
\hline
 &  $d$ & $q$ &  $e^{\frac{-a}{\tau}}$ & $d q$ & $d e^{\frac{-a}{\tau}}$  & $q e^{\frac{-a}{\tau}}$ & $d q e^{\frac{-a}{\tau}}$ \\
\hline
$d$ & - & 0.22 & 0.30 & 0.43 & 0.18 & 0.22 & 0.19 \\
\hline
$q$ & 0.78 & - & 0.38 & 0.76 & 0.41 & 0.23 & 0.40 \\
\hline
$e^{\frac{-a}{\tau}}$ & 0.70 & 0.62 & - & 0.69 & 0.54 & 0.40 & 0.53 \\
\hline
$d q$ & 0.57 & 0.24 & 0.31 & - & 0.24 & 0.23 & 0.17 \\
\hline
$d e^{\frac{-a}{\tau}}$ & 0.82 & 0.59 & 0.44 & 0.76 & - & 0.39 & 0.43 \\
\hline
$q e^{\frac{-a}{\tau}}$ & 0.78 & 0.77 & 0.60 & 0.77 & 0.61 & - & 0.62 \\
\hline
$d q e^{\frac{-a}{\tau}}$ & 0.81 & 0.60 & 0.47 & 0.83 & 0.57 & 0.38 & -  \\
\hline
\end{tabular}
\end{center}
\end{table}

\begin{figure}
        \centering
        \includegraphics[width=0.68\textwidth]{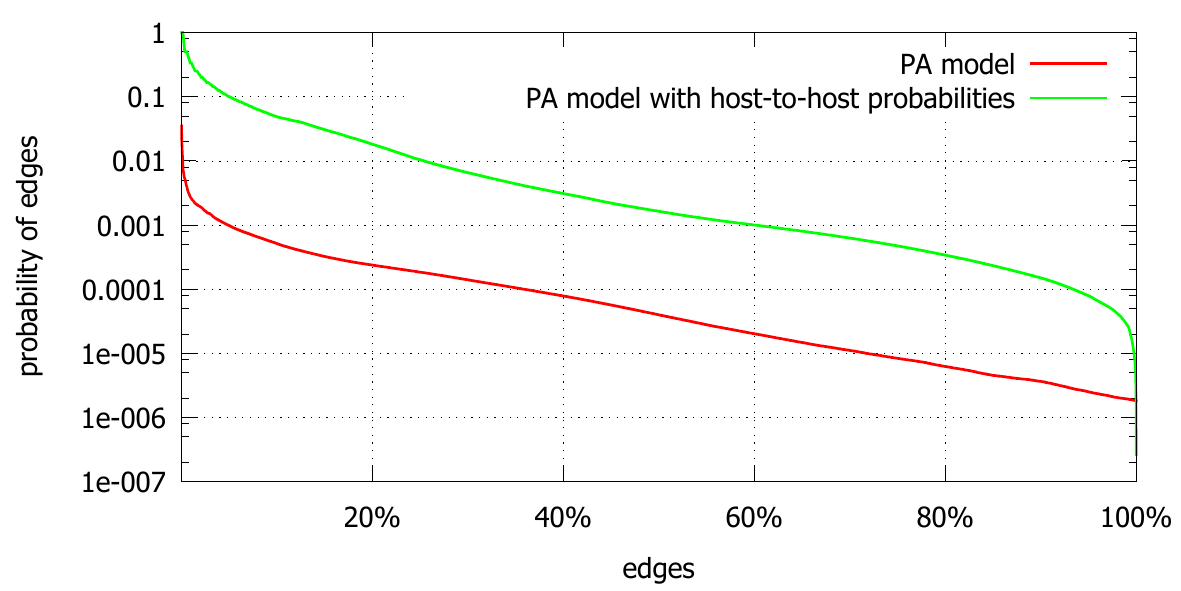}
        \caption{The influence of host-to-host probabilities, e.g., on preferential attachment model (PA)}
        \label{fig:nohost}
\end{figure}

Then, for each model, we sorted edges' probabilities in decreasing
order on Figure \ref{fig:edges}. Furthermore, in order to more clearly
visualize the differences between models, we normalized the
probability of each edge in all models by dividing it by the
corresponding probability in the sorted order of the preferential
attachment model (see Figure~\ref{fig:edges_rel}). One can see that
the model with $f_{\tau} = q e^{\frac{-a}{\tau}}$ again shows the best
result in our tests. This means that in the Media Web the probability
for a post to be cited is determined, most likely, by its quality
rather than by its current popularity (i.e., incoming
degree). Finally, the importance of host-to-host probabilities
$\rho_{ij}$ can be illustrated by Figure~\ref{fig:nohost}.

\section{Conclusion}\label{Conclusion}

In this paper, we presented a detailed study of the Media Web. We proposed
a new class of models for the appearance of new media content on the Web where different
attractiveness functions of nodes are possible including ones
taken from well-known preferential attachment and fitness models, but
also new ones accounting for specificities of this part of the
Web. Our new models are based on the observation that media pages tend to connect
with other media pages of similar age.

We analyzed these models theoretically and empirically using publicly
available data and show which ones realistically predict both the
incoming degree distribution and the so-called \textit{recency property} of the
Media Web, something that existing models did not do well.

Finally we
compared these models by estimating the likelihood of the real-world
link graph from our data set given each model and obtained that new models
we introduce, with a recency
factor, are significantly more likely than previously proposed
ones.  One of the most surprising results is that in the Media Web the
probability for a post to be cited is determined, most likely, by its
quality rather than by its current popularity.




\bibliography{references}
\bibliographystyle{splncs03}


\newpage

\section*{Appendix: Proof of Theorem~\ref{Theorem2}}

To analyze the behavior of $e(T)$, we need to estimate the average attractiveness of all
  media pages created in the last $T$ seconds at time $t$ at a host $k$:
$$\mathcal{W}_k(T, t) = \mathbb{E}
       \sum_{\begin{subarray}{c}
              p\in\Omega(H_k) \\ |t-t_p|<T
             \end{subarray}} f_{\tau_k}(d_p(q_p, t, t_p), q_p, t - t_p).$$
We will show that if ${t>T}$, then this function does not depend on $t$.

We can analyze the function $\mathcal{W}_k(T, t)$ using the technique we used in Section~\ref{Degree}. Consider the case $f_{\tau_k}(d, q, a) = q \cdot d
\cdot e^{-\frac{a}{\tau_k}}$:
\begin{multline*}
\mathcal{W}_k(T, t) = \\
 =\int_0^{\infty} \left( \int_{t-T}^t \lambda_k  q \rho_k(q) d(q, t, x) \cdot e^{-\frac{t-x}{\tau_k}} d x \right) d q = \\
=  \int_0^{\infty}
  \left( \int_{t-T}^t  \lambda_k q \rho_k(q)
     e^{ \frac{N_k \tau_k q}{W_k} \left(1 - e^{\frac{x-t}{\tau_k}} \right) }
     \cdot e^{\frac{x-t}{\tau_k}} d x
  \right) d q = \\
= \frac{\lambda_k W_k}{N_k} \int_0^{\infty} \left( e^{ \frac{N_k \tau_k q}{W_k} \left(1 - e^{\frac{-T}{\tau_k}}\right) } - 1 \right) \rho_k(q) d q\,  .
\end{multline*}

We proved that $\mathcal{W}_k(T,t)$ does not depend on $t$ and will use the notation $\mathcal{W}_k(T) = \mathcal{W}_k(T,t)$ from now on.
Also
\begin{multline*}
W_k - \mathcal{W}_k(T) = \\
= \frac{\lambda_k W_k}{N_k} \int_0^{\infty}   \left(1 - e^{ -\frac{N_k \tau_k q}{W_k} e^{\frac{-T}{\tau_k}}}\right) e^{ \frac{N_k \tau_k q}{W_k}} \rho_k(q) d q \sim \\
\sim
\frac{\lambda_k W_k}{N_k} e^{\frac{-T}{\tau_k}} \int_0^{\infty}   \frac{N_k \tau_k q}{W_k} e^{ \frac{N \tau_k q}{W}} \rho_k(q) d q \sim C_k e^{\frac{-T}{\tau_k}},
\end{multline*}
where the constants $C_k$ do not depend on $T$.

Note that the portion of links which point to the host $H_k$ and have the age difference less than $T$ is $\frac{W_k-\mathcal{W}_k(T)}{W_k}$. Thus, using $N_k$ which is the average rate of new links pointing to host $H_k$ appearance (see Section~\ref{Degree}) we can write the following
equation for $e(T)$:

$$
e(T) = \sum_k N_k \frac{W_k-\mathcal{W}_k(T)}{W_k} \sim \sum_k \frac{N_k C_k}{W_k} e^{\frac{-T}{\tau_k}}
$$









The same analysis can be made for the case $f_{\tau_k}(d, q, a) = q \cdot e^{-\frac{a}{\tau_k}}$. In this case we get:
\begin{multline*}
\mathcal{W}_k(T)
 = \int_0^{\infty} \left( \int_{t-T}^t \lambda_k  q \rho_k(q) \cdot e^{-\frac{t-x}{\tau_k}} d x \right) d q 
=  \lambda_k \tau_k \left(1 - e^{-\frac{T}{\tau_k}}\right) \mathbb{E}_{p \in \Omega(H_k)}  q_p \, ,
\end{multline*}
and further reasonings are the same.

\end{document}